\documentclass[
  aps,
  prb,
  superscriptaddress,
  floatfix,
  twocolumn,
  secnumarabic,
  amssymb,
  amsmath]{revtex4}

\usepackage{dcolumn}
\usepackage{bm}
\usepackage{hyperref}
\usepackage[utf8]{inputenc} 
\usepackage[pdftex]{graphicx}
\usepackage{tabularx}
\usepackage{subfigure}
\usepackage{amssymb}
\usepackage{enumerate}
\usepackage{array}
\usepackage{siunitx}

\begin{document}

\title[Spin Hall magnetoresistance in antiferromagnetic insulators]{Spin Hall magnetoresistance in antiferromagnetic insulators}

\author{Stephan~Geprägs}
  \email{stephan.gepraegs@wmi.badw.de}
	\affiliation{Walther-Mei{\ss}ner-Institut, Bayerische Akademie der Wissenschaften, 85748 Garching, Germany}
\author{Matthias~Opel}
  \email{matthias.opel@wmi.badw.de}
  \affiliation{Walther-Mei{\ss}ner-Institut, Bayerische Akademie der Wissenschaften, 85748 Garching, Germany}
\author{Johanna~Fischer}%
  \altaffiliation[Present address: ]{Unit\'{e} Mixte de Physique, CNRS, Thales, Universit\'{e} Paris-Sud, Universit\'{e} Paris-Saclay, 91767 Palaiseau, France}
  \affiliation{Walther-Mei{\ss}ner-Institut, Bayerische Akademie der Wissenschaften, 85748 Garching, Germany}
  \affiliation{Physik-Department, Technische Universit\"{a}t M\"{u}nchen, 85748 Garching, Germany}
\author{Olena Gomonay}
   \affiliation{Institut f\"{u}r Physik, Johannes Gutenberg Universit\"{a}t Mainz, 55128 Mainz, Germany}	
\author{Philipp~Schwenke}
  \affiliation{Walther-Mei{\ss}ner-Institut, Bayerische Akademie der Wissenschaften, 85748 Garching, Germany}
	\affiliation{Physik-Department, Technische Universit\"{a}t M\"{u}nchen, 85748 Garching, Germany}
\author{Matthias~Althammer}
  \affiliation{Walther-Mei{\ss}ner-Institut, Bayerische Akademie der Wissenschaften, 85748 Garching, Germany}
  \affiliation{Physik-Department, Technische Universit\"{a}t M\"{u}nchen, 85748 Garching, Germany}
\author{Hans~Huebl}
  \affiliation{Walther-Mei{\ss}ner-Institut, Bayerische Akademie der Wissenschaften, 85748 Garching, Germany}
  \affiliation{Physik-Department, Technische Universit\"{a}t M\"{u}nchen, 85748 Garching, Germany}
  \affiliation{Munich Center for Quantum Science and Technology (MCQST), 80799 Munich, Germany}
\author{Rudolf~Gross}
  \affiliation{Walther-Mei{\ss}ner-Institut, Bayerische Akademie der Wissenschaften, 85748 Garching, Germany}
  \affiliation{Physik-Department, Technische Universit\"{a}t M\"{u}nchen, 85748 Garching, Germany}
  \affiliation{Munich Center for Quantum Science and Technology (MCQST), 80799 Munich, Germany}

\date{\today}

\begin{abstract}
    Antiferromagnetic materials promise improved performance for spintronic applications, as they are robust against external magnetic field perturbations and allow for faster magnetization dynamics compared to ferromagnets. The direct observation of the antiferromagnetic state, however, is challenging due to the absence of a macroscopic magnetization. Here, we show that the spin Hall magnetoresistance (SMR) is a versatile tool to probe the antiferromagnetic spin structure via simple electrical transport experiments by investigating the easy-plane antiferromagnetic insulators $\alpha$-Fe$_2$O$_3$ (hematite) and NiO in bilayer heterostructures with a Pt heavy metal top electrode. While rotating an external magnetic field in three orthogonal planes, we record the longitudinal and the transverse resistivities of Pt and observe characteristic resistivity modulations consistent with the SMR effect. We analyze both their amplitude and phase and compare the data to the results from a prototypical collinear ferrimagnetic Y$_3$Fe$_5$O$_{12}$/Pt bilayer. The observed magnetic field dependence is explained in a comprehensive model, based on two magnetic sublattices and taking into account magnetic field-induced modifications of the domain structure. Our results show that the SMR allows us to understand the spin configuration and to investigate magnetoelastic effects in antiferromagnetic multi-domain materials. Furthermore, in $\alpha$-Fe$_2$O$_3$/Pt bilayers, we find an unexpectedly large SMR amplitude of $2.5 \times 10^{-3}$, twice as high as for prototype Y$_3$Fe$_5$O$_{12}$/Pt bilayers, making the system particularly interesting for room-temperature antiferromagnetic spintronic applications.
\end{abstract}

\maketitle

\section{
  \label{sec:Intro}
  Introduction
  }

Despite lacking a macroscopic magnetization, antiferromagnetic (AF) materials have moved into the focus of spintronics research.\cite{Marti:2014, Jungwirth:2016, Baltz:2018} This material class brings along two important advantages compared to ferromagnets: (i) better scalability and improved robustness against magnetic field perturbations,\cite{Marti:2014, Jungwirth:2016} and (ii) orders of magnitudes faster dynamics and switching times.\cite{Satoh:2014, Olejnik:2018} From an application perspective, however, it is evident that their vanishing macroscopic magnetization and stray fields call for new methods for magnetization control and read-out. To this end, spin currents\cite{Althammer:2018a} were shown to interact with the individual magnetic sublattices in ferrimagnets\cite{Ando:2008, Miron:2011, Jia:2011, Liu:2012} and antiferromagnets\cite{Wadley:2016, Matalla-Wagner:2019, Baldrati:2019} via spin transfer torques. Spin currents can be created via the spin pumping effect in radio-frequency magnetoelectric fields,\cite{Tserkovnyak:2002B, Mosendz:2010, Ando:2010, Heinrich:2011, Hahn:2013L, Jiao:2013} the spincaloric effect in thermal gradients,\cite{Uchida:2008, Xiao:2010, Bauer:2012} or the spin Hall effect in metals with large spin-orbit interaction.\cite{Hirsch:1999} For the realization of spin current devices, electrically insulating oxide materials are beneficial since they prevent spurious charge transport effects.\cite{Coll:2019} Fortunately, most antiferromagnets are electrical insulators, however, they exhibit a more complex spin texture than ferrimagnets.\cite{Gomonay:2018} Recently, we showed that the spin Hall magnetoresistance (SMR) allows to obtain valuable information on the spin texture via straightforward electrical transport measurements.\cite{Nakayama:2013, Althammer:2013, Chen:2013} Here, we review our work on the SMR in bilayers of Pt and antiferromagnets,\cite{Fischer:2018, Fischer:2020} complemented by numerous additional data sets for out-of-plane rotations of the magnetic field. We show the similarities and the striking differences of the angular-dependent magnetotransport data from the established situation in ferrimagnetic materials.\cite{Nakayama:2013, Althammer:2013}

\section{Theory
  \label{sec:SMR}}

A well known manifestation of the spin current physics is the dependence of the resistivity of a metallic thin film with large spin Hall effect on the directions of the sublattice magnetizations in an adjacent insulating magnetic material. This effect is denoted as spin Hall magnetoresistance (SMR).\cite{Nakayama:2013, Althammer:2013, Chen:2013, Vlietstra:2013, Hahn:2013, Marmion:2014, Meyer:2014, Isasa:2014, Aldosary:2016, Ganzhorn:2016, Hoogeboom:2017, Fischer:2018, Baldrati:2018, Ji:2018, DuttaGupta:2018, Althammer:2019, Cheng:2019, Fischer:2020} It is based on an interfacial exchange of angular momentum from the sublattice magnetizations to the conduction electrons of the heavy metal and has to be distinguished from a static spin polarization due to magnetic proximity effects.\cite{Geprags:2012, Liang:2016, Vasili:2018}

\subsection{
  \label{sec:SMR-Phenom}
  The spin Hall magnetoresistance
  }

A heavy metal with a large spin-orbit interaction gives rise to the spin Hall effect,\cite{Dyakonov:1971,Hirsch:1999,Kato:2004} which manifests itself in a spin current density $\mathbf{J}_s$ perpendicular to the charge current density $\mathbf{J}$ and the spin polarization $\boldsymbol{\sigma$} of the conduction electrons leading to a spin accumulation at the edges of the heavy metal. This gradient of the spin chemical potential results in a diffusive spin current backflow, compensating $\mathbf{J}_s$. Putting a magnetic insulating material with a local magnetization $\mathbf{M}$ adjacent to this heavy metal, an interfacial spin mixing between the magnetic insulator and the heavy metal occurs, which leads to a spin transfer torque on $\mathbf{M}$, if $\boldsymbol{\sigma}$ and $\mathbf{M}$ are not collinear. The transfer of spin angular momentum across the interface corresponds to a finite spin current density and leads to a reduction of the spin accumulation. It represents an additional loss channel, leading to an increase of the electrical resistivity of the heavy metal. The resistivity is expected to be maximum (minimum) for $\mathbf{M} \perp \boldsymbol{\sigma}$ ($\mathbf{M} || \boldsymbol{\sigma}$). This magnetoresistance effect, originating from the concerted action of the direct and inverse spin Hall effects, is known as ``spin Hall magnetoresistance'' (SMR).\cite{Nakayama:2013} 

The SMR was first demonstrated in ferrimagnetic Y$_3$Fe$_5$O$_{12}$/Pt heterostructures\cite{Nakayama:2013} and theoretically described in general for multilayer samples.\cite{Chen:2013} During the past years, the SMR has been measured in numerous experiments in various material systems: In bilayers using collinear ferrimagnets such as Y$_3$Fe$_5$O$_{12}$/Pt\cite{Nakayama:2013, Vlietstra:2013, Althammer:2013, Hahn:2013, Marmion:2014, Meyer:2014, Aldosary:2016}, Y$_3$Fe$_5$O$_{12}$/Ta\cite{Hahn:2013}, Fe$_3$O$_4$/Pt\cite{Althammer:2013}, NiFe$_2$O$_4$/Pt\cite{Althammer:2013,Althammer:2019}, CoFe$_2$O$_4$/Pt\cite{Isasa:2014}, and $\gamma$-Fe$_2$O$_3$/Pt\cite{Dong:2019}, as well as in bilayers using antiferromagnetic materials such as NiO/Pt\cite{Hoogeboom:2017, Fischer:2018, Baldrati:2018}, Cr$_2$O$_3$/Ta\cite{Ji:2018}, or $\alpha$-Fe$_2$O$_3$/Pt\cite{Cheng:2019, Lebrun:2019, Fischer:2020}. Furthermore, the SMR was successfully applied to reveal the canted spin structure in the compensated garnet YGd$_2$Fe$_4$InO$_{12}$\cite{Ganzhorn:2016} or the spiral texture in the skyrmion compound Cu$_2$OSeO$_3$\cite{Aqeel:2016}. A finite SMR effect was reported even for non-crystalline, paramagnetic Y$_3$Fe$_5$O$_{12}$/Pt\cite{Lammel:2019} bilayers and in antiferromagnetic Cr$_2$O$_3$/Pt heterostructures above the N\'{e}el temperature, in the paramagnetic state of Cr$_2$O$_3$.\cite{Schlitz:2018} 

In all these examples, the magnetic insulators are composed of at least two magnetic sublattices. In the following we thus discuss the SMR in bilayers consisting of a heavy metal deposited on a magnetically ordered, electrical insulator with two magnetic sublattices.

\subsection{
  \label{sec:SMR-versusM}
  Spin Hall magnetoresistance in two-sublattice magnetic insulators
  }
	
\begin{figure}
  \includegraphics[width=\columnwidth]{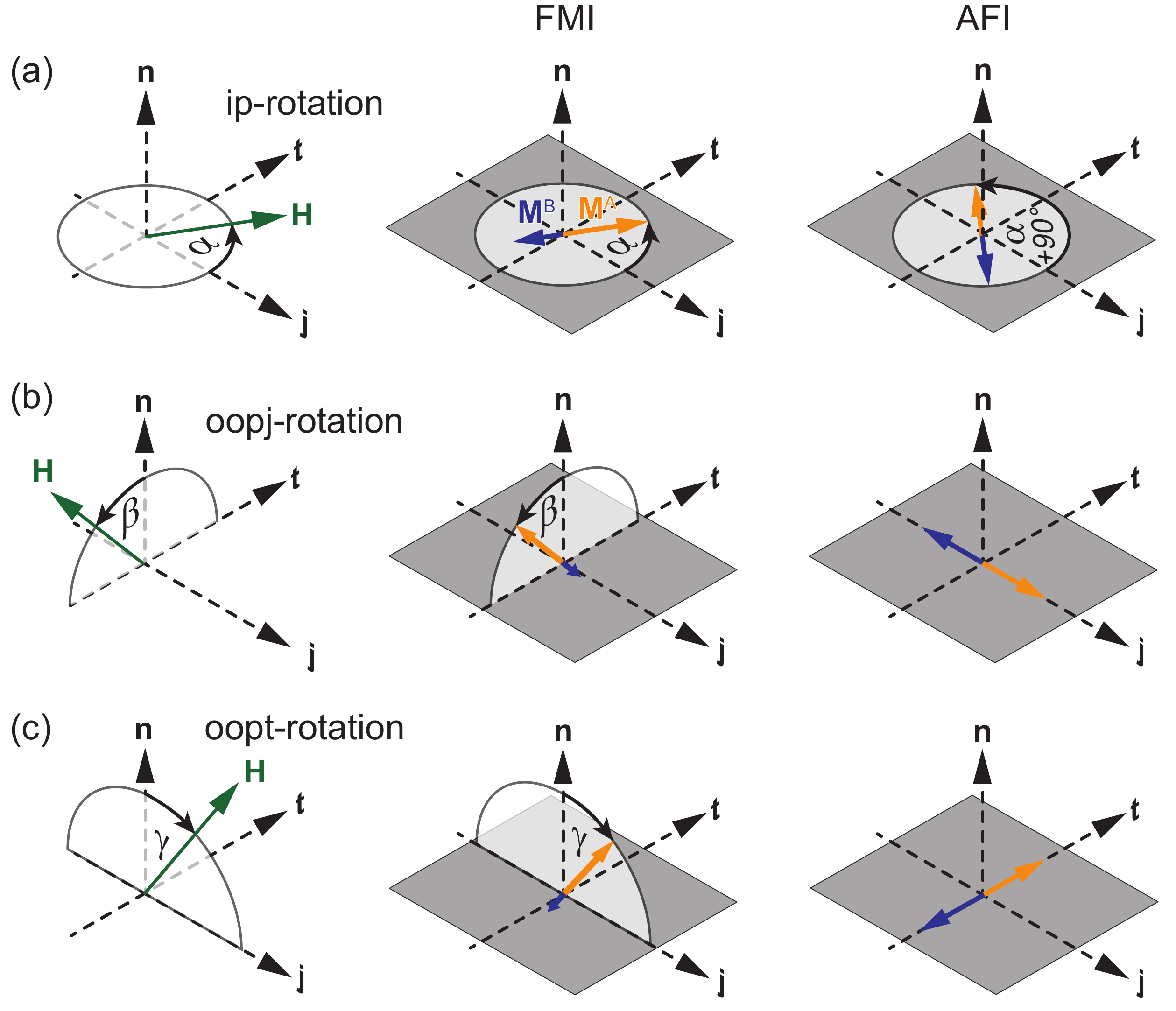}
  \caption{
    \label{fig:alpha-phi}
    Angle-dependence of the sublattice magnetizations $\mathbf{M}^A$ and $\mathbf{M}^B$ (orange and blue arrows) of a two-sublattice ferrimagnetic insulator (FMI) with $|\mathbf{M}^A| > |\mathbf{M}^B|$ and an easy-plane insulating antiferromagnet (AFI) with $|\mathbf{M}^A| = |\mathbf{M}^B|$ when rotating an external magnetic field $\mathbf{H}$ (green arrow) in different planes: (a) in the $\mathbf{j}$-$\mathbf{t}$ plane with angle $\alpha$ (ip-rotation), (b) in the $\mathbf{t}$-$\mathbf{n}$ plane with angle $\beta$ (oopj-rotation), and (c) in the $\mathbf{n}$-$\mathbf{j}$ plane with angle $\gamma$ (oopt-rotation). In this simple picture neglecting any anisotropy and domain effects as well as spin canting, the net magnetization of the ferrimagnet $\mathbf{M}=\mathbf{M}^A+\mathbf{M}^B$ follows the magnetic field $\mathbf{H}$, while in antiferromagnets $\mathbf{M}^A$ and $\mathbf{M}^B$ stay within the easy-plane (grey area).  
    }
\end{figure}

Due to the SMR theory, the modulation of the resistivity of a heavy metal layer (e.g.~Pt) depends on the direction $\mathbf{m}^\mathrm{X}=\mathbf{M}^\mathrm{X}/M_\mathrm{s}^\mathrm{X}$ of the sublattice magnetizations $\mathbf{M}^\mathrm{X}$ with the saturation magnetization $M_\mathrm{s}^\mathrm{X}$ of the adjacent uniformly magnetically ordered insulator with two magnetic sublattices $X=A,B$.\cite{Chen:2013, Ganzhorn:2016} The longitudinal resistivity $\rho_\mathrm{long}$ of a polycrystalline heavy metal measured along the current direction $\mathbf{j}=\mathbf{J}/J$ is given by\cite{Fischer:2018,Ganzhorn:2016, Chen:2013}
\begin{align}
\rho_{\mathrm{long}}&= \rho_{0}+\frac{1}{2}\sum_{X=A}^B \rho_1^X \left[ 1 - \left(\mathbf{m}^X \cdot \mathbf{t}\right)^{2}\right] \nonumber \\
	&= \rho_{0}+\frac{1}{2}\sum_{X=A}^B \rho_1^X \left[1 - \left(m_t^X\right)^2 \right]
	\; .
	\label{eq:SMR-eq1}
\end{align}
Here, $\rho_{0}$ is approximately equal to the normal resistivity of the metallic layer,\cite{Chen:2013} $\rho_1^X$ represents the SMR coefficient for the magnetic sublattice $X=A,B$ with $\rho_1^X \ll \rho_0$, and $m_t^X = \mathbf{m}^X \cdot \mathbf{t}$ denotes the projection of $\mathbf{m}^X$ on the transverse direction $\mathbf{t}$ (perpendicular to $\mathbf{j}$ in the $\mathbf{j}$-$\mathbf{t}$-interface plane, see Fig.~\ref{fig:alpha-phi}).

From a similar consideration, the transverse resistivity $\rho_\mathrm{trans}$ is given by \cite{Chen:2013, Ganzhorn:2016, Althammer:2013}
\begin{align}
  \rho_{\mathrm{trans}}	= \frac{1}{2}\sum_{X=A}^B \left[ \rho_3^X \, m_j^X \, m_t^X + \rho_2^X \, m_n^X \right]
	\label{eq:SMR-eq1a}
\end{align}
with the transverse SMR coefficient $\rho_3^X \ll \rho_0$ and an anomalous Hall-effect-type resistivity coefficient $\rho_2^X$. $m_j^X$, $m_n^X$ are the projections of $\mathbf{m}^X$ on the directions of the current density $\mathbf{j}$ and the surface normal $\mathbf{n}$, respectively (see~Fig.~\ref{fig:alpha-phi}). In the following, we will only discuss the SMR coefficients $\rho_1^X$ and $\rho_3^X$, which should be equal in the framework of the SMR theory.\cite{Althammer:2013}   

We now consider ferrimagnetic insulators with two antiferromagnetically coupled magnetic sublattices $\mathbf{M}^A$, $\mathbf{M}^B$ with $|\mathbf{M}^A| > |\mathbf{M}^B|$ (e.g.~Y$_3$Fe$_5$O$_{12}$, Fe$_3$O$_{4}$) and easy-plane antiferromagnets with $|\mathbf{M}^A| = |\mathbf{M}^B|$ (e.g.~NiO) and apply a rotating external magnetic field $\mathbf{H}$ in the $\mathbf{j}$-$\mathbf{t}$ plane (ip-rotation with angle $\alpha$), the $\mathbf{t}$-$\mathbf{n}$ plane (oopj-rotation with angle $\beta$), or the $\mathbf{n}$-$\mathbf{j}$ plane (oopt-rotation with angle $\gamma$), respectively (see Fig.~\ref{fig:alpha-phi}). For simplicity, we first neglect any magnetic anisotropy or domain effects within the easy-plane of the antiferromagnetic insulator as well as for the ferrimagnetic insulator. Furthermore, we do not consider any spin canting of the magnetic sublattices. This is valid for magnetic-field energies larger than the magnetic anisotropy energies and much smaller than the exchange interactions between the two magnetic sublattices.   

In a \textit{ferrimagnet} with the net magnetization $\mathbf{M}=\mathbf{M}^A +\mathbf{M}^B$ following the magnetic field (see Fig.~\ref{fig:alpha-phi}), Equations~(\ref{eq:SMR-eq1}) and (\ref{eq:SMR-eq1a}) simplify to  
\begin{eqnarray}
\mathrm{ip} &:&\;\rho_\mathrm{long}^\mathrm{FMI}(\alpha) \; = \rho_{0} + \frac{\rho_1}{2} \left[1 + \cos2\alpha\right] \nonumber \\
            & &\;\rho_\mathrm{trans}^\mathrm{FMI}(\alpha)	\;=\; \frac{\rho_3}{2} \sin2\alpha \nonumber \\
\mathrm{oopj} &:&\;\rho_\mathrm{long}^\mathrm{FMI}(\beta) \; = \rho_{0} + \frac{\rho_1}{2} \left[1 + \cos2\beta\right] \nonumber \\
\mathrm{oopt} &:&\;\rho_\mathrm{long}^\mathrm{FMI}(\gamma) \; = \rho_{0} + \rho_1			
     \;\; .
		\label{eq:SMR-FMI}
\end{eqnarray}
Here, we assume that the SMR coefficients of the two magnetic sublattices are equal: $\rho_1 = \rho_1^A = \rho_1^B$ and $\rho_3 = \rho_3^A = \rho_3^B$. Equations~(\ref{eq:SMR-FMI}) reveal the characteristic variation of the resistivity in bilayers based on ferro-/ferrimagnetic insulator with sinusoidal oscillations of the longitudinal resistivity $\rho_\mathrm{long}$ for ip- and oopj-magnetic field rotations and a constant value $\rho_\mathrm{long}=\rho_{0} + \rho_1$ for oopt-rotations. For the transverse resistivity $\rho_\mathrm{trans}$, a resistivity modulation originating from SMR is only visible for in-plane rotations, while in out-of-plane rotations (oopj- and oopt-rotations), $\rho_\mathrm{trans}$ is dominated by the ordinary Hall effect and will not be discussed in this paper.\cite{Althammer:2013}        

In easy-plane \textit{antiferromagnets}, however, the situation is more complex. For magnetic field rotations in the easy plane, the magnetic sublattices $A$ and $B$ rotate perpendicular to $\mathbf{H}$ within the $\mathbf{j}$-$\mathbf{t}$ plane (cf.~Fig.~\ref{fig:alpha-phi} (a)).\cite{Fischer:2018} In out-of-plane rotations, however, the sublattice magnetizations $\mathbf{M}^A$ and $\mathbf{M}^B$ stay within the easy plane resulting in constant resistivity values (cf.~Fig.~\ref{fig:alpha-phi} (b),(c)). From Eqs.~(\ref{eq:SMR-eq1}) and (\ref{eq:SMR-eq1a}), we get
\begin{eqnarray}
\mathrm{ip} &:&\;\rho_\mathrm{long}^\mathrm{AFI}(\alpha) \; = \rho_{0} + \frac{\rho_1}{2} \left[1 - \cos2\alpha\right] \nonumber \\
            & &\;\rho_\mathrm{trans}^\mathrm{AFI}(\alpha)	\;=\; -\frac{\rho_3}{2} \sin2\alpha \nonumber \\
\mathrm{oopj} &:&\;\rho_\mathrm{long}^\mathrm{AFI}(\beta) \; = \rho_{0} + \rho_1 \nonumber \\
\mathrm{oopt} &:&\;\rho_\mathrm{long}^\mathrm{AFI}(\gamma) \; = \rho_{0} 	
     \;\; ,
		\label{eq:SMR-AFI}
\end{eqnarray}
with $\rho_1 = \rho_1^A = \rho_1^B$ and $\rho_3 = \rho_3^A = \rho_3^B$. Because of the minus signs in the expressions for $\rho_\mathrm{long}^\mathrm{AFI}(\alpha)$ and $\rho_\mathrm{trans}^\mathrm{AFI}(\alpha)$, the SMR in antiferromagnets is also referred to as \textit{negative} spin Hall magnetoresistance.

\subsection{
  \label{sec:SMR-multidomain}
  Spin Hall magnetoresistance in multi-domain magnetic insulators
  }

In reality, the antiferromagnetic insulator forms magnetic domains within the easy plane with different orientations of the sublattice magnetizations.\cite{Fischer:2018} This can be taken into account by introducing fractions $\xi_k$ of the domains $k$ with $\sum_{k} \xi_k = 1$. Neglecting any contributions from domain walls, we get\cite{Fischer:2018}
\begin{eqnarray}
	\rho_\mathrm{long}^\mathrm{AFI} &=&\rho_0+\rho_1\sum_{k}\xi_k \left[1 - \left(\ell_{t}^{(k)} \right)^2\right], \nonumber\\
  \rho_\mathrm{trans}^\mathrm{AFI} &=&\rho_3\sum_{k}\xi_k \; \ell_{j}^{(k)} \ell_{t}^{(k)}
\, .
 \label{eq:SMR-AFI-domains}
	\end{eqnarray}
$\ell_j^{(k)}$ and $\ell_t^{(k)}$ represent the projections of the unit vector $\boldsymbol{\ell}^{(k)} = (\mathbf{m}^{A,(k)} - \mathbf{m}^{B,(k)})/2$ of the domain $k$ on $\mathbf{j}$ and $\mathbf{t}$, respectively. In Eqs.~(\ref{eq:SMR-AFI-domains}), we average over the heavy-metal resistance contributions from the individual antiferromagnetic domains taking into account the relative fraction of each domain $\xi_k$. Therefore, for the calculation of the longitudinal and transverse resistivities, a detailed knowledge of the domain structure in the presence of an applied magnetic field is required. For magnetic fields $\mathbf{H}$ oriented (anti-)parallel to the surface normal $\mathbf{n}$, however, we can assume an equal distribution of the domains within the easy plane of the antiferromagnetic insulator. We therefore expect $\rho_\mathrm{long}^\mathrm{AFI}$ to be         
\begin{eqnarray}
\rho_\mathrm{long}^\mathrm{AFI}(\beta=\gamma=0^\circ) &=& \rho_{0} + \frac{\rho_1}{2} 			
     \;\; .
		\label{eq:SMR-AFI-oop}
\end{eqnarray}

\section{
  \label{sec:expt}
  Experimental
  }
To validate the above model, we will discuss experimental results of the SMR in heavy metal/magnetic insulator bilayers with Y$_3$Fe$_5$O$_{12}$ (yttrium iron garnet, YIG) as the ferrimagnetic material and NiO as well as $\alpha$-Fe$_2$O$_3$ (hematite) as easy-plane antiferromagnetic insulators. For the heavy metal, we choose Pt as the prototype material. To this end, we fabricate $\alpha$-Fe$_2$O$_3$/Pt, NiO/Pt and Y$_3$Fe$_5$O$_{12}$/Pt thin film bilayer samples. Using photolithography, we pattern the bilayers into Hall bar-shaped mesa structures and subsequently investigate their magnetotransport properties for various rotations of the applied magnetic fields. 

\subsection{
  \label{sec:expt-materials}
  Magnetic materials
  }
The antiferromagnetic insulator $\alpha$-Fe$_2$O$_3$ crystallizes in a hexagonal (rhombohedral) structure and exhibits a N\'{e}el temperature of $T_\mathrm{N}=953$\,K with a spin reorientation (``Morin'' transition) at $T_\mathrm{M} \approx 263$\,K. \cite{Morin:1950} For $T_\mathrm{M} < T < T_\mathrm{N}$, i.e.~at room temperature, the spin structure of $\alpha$-Fe$_2$O$_3$ is given by the $S=5/2$ spins of the Fe$^{3+}$ ions, which are ordered in two antiferromagnetically coupled sublattices perpendicular to the $[0001]$-direction in the basal plane of the hexagonal structure.\cite{Shull:1951} Within this magnetically easy plane the anisotropic Dzyaloshinskii-Moriya-Interaction (DMI) leads to a finite canting of the two magnetic sublattices towards each other, resulting in a weak net magnetization of $M = 2.5$\,kA/m at room temperature.\cite{Coey:book}

NiO represents a prototypical biaxial antiferromagnetic insulator with a N\'{e}el temperature of $T_\mathrm{N} = 523$\,K,\cite{Srinivasan:1984} crystallizing in the cubic NaCl structure. At room temperature, the Ni$^{2+}$ spins align ferromagnetically along the cubic $\langle11\overline{2}\rangle$ directions within the easy \{111\} planes and antiferromagnetically between neighboring \{111\} planes.\cite{Roth:1958,Hutchings:1972} Both $\alpha$-Fe$_2$O$_3$\cite{Nathans:1964, Marmeggi:1977} and NiO\cite{Roth:1958,Hutchings:1972} display three antiferromagnetic domains rotated by $120^\circ$ with respect to each other and a domain population dependent on the direction and magnitude of the external magnetic field.\cite{Fischer:2018, Fischer:2020,Baldrati:2018}

Y$_3$Fe$_5$O$_{12}$ crystallizes in the cubic garnet structure and exhibits a Curie temperature of $T_\mathrm{C} = 560$\,K.\cite{Coey:book} The five Fe$^{3+}$ ions per formula unit occupy two octahedrally ($a$-sites) and three tetrahedrally ($d$-sites) coordinated lattice sites. They form two opposite magnetic sublattices with non-compensating sublattice magnetizations, resulting in a ferrimagnetic order with a saturation magnetization of $M_s = 143$\,kA/m at room temperature.\cite{Coey:book}

\subsection{
  \label{sec:expt-samples}
  Sample fabrication and characterization
  }

\begin{figure}
  \includegraphics[width=0.9\columnwidth]{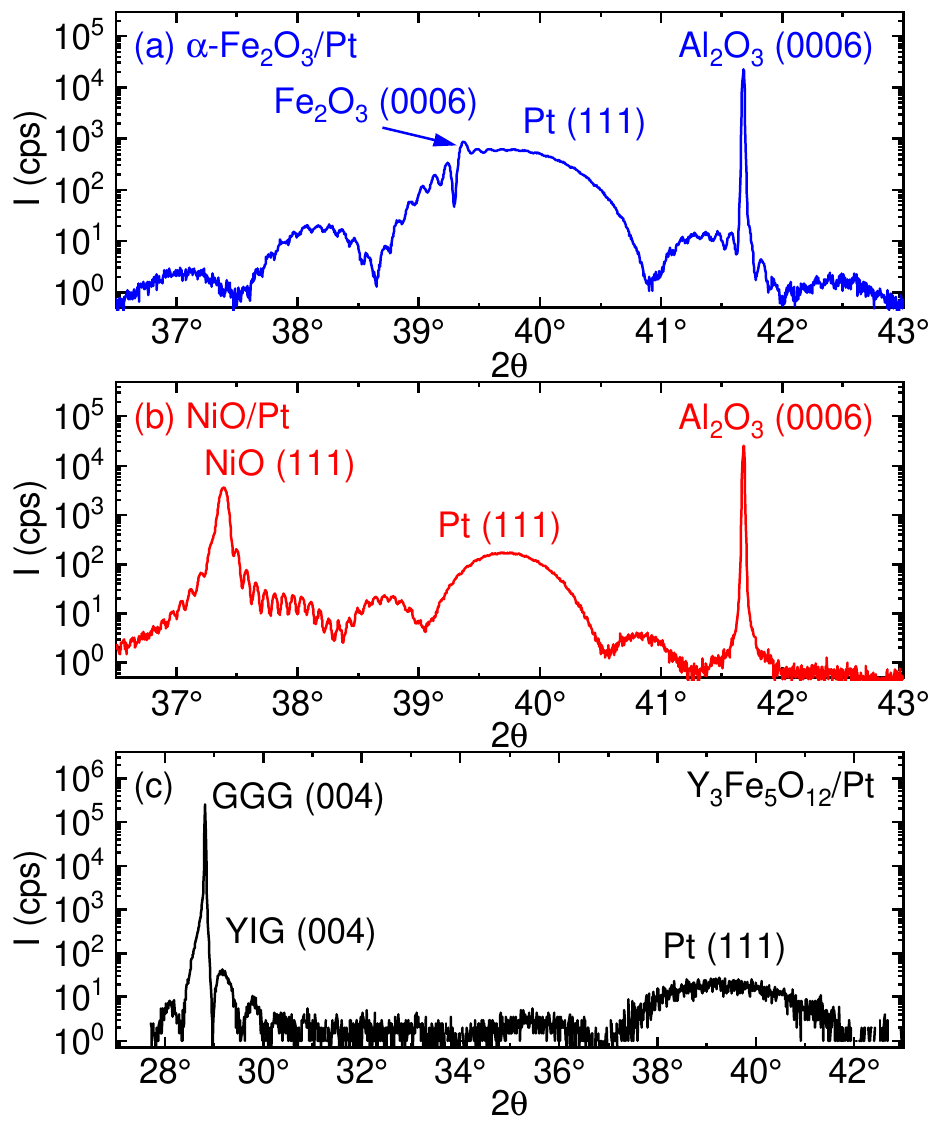}
  \caption{
    \label{fig:XRD}
    High-resolution X-ray diffraction of (a) a $\alpha$-Fe$_2$O$_3$/Pt (96.1\,nm/8.4\,nm), (b) a NiO/Pt (92.2\,nm/12.9\,nm), and (c) a Y$_3$Fe$_5$O$_{12}$ (YIG)/ Pt (20.1\,nm/3.3\,nm) bilayer on (0001)-oriented Al$_{2}$O$_{3}$ and (001)-oriented Gd$_3$Ga$_5$O$_{12}$ (GGG) substrates, respectively. Clearly, finite thickness fringes around the symmetric Fe$_2$O$_3$(0006), NiO(111), and YIG(004) reflections as well as the Pt(111) reflections are visible, indicating a coherent growth with low surface roughness.
    }
\end{figure}

The magnetically ordered oxide thin films are fabricated by pulsed-laser deposition from stoichiometric targets in an oxygen atmosphere on single crystalline substrates, utilizing a KrF excimer laser with a wavelength of 248\,nm.\cite{Opel:2014} The (0001)-oriented $\alpha$-Fe$_2$O$_3$ as well as the (111)-oriented NiO thin films are deposited on single-crystalline, (0001)-oriented Al$_{2}$O$_{3}$ substrates with a laser fluence of $2.5\,\mathrm{J/cm^2}$ and a repetition rate of $f=2$\,Hz. Best structural and magnetic properties of the $\alpha$-Fe$_2$O$_3$ and NiO films were obtained using a substrate temperature of $T_\mathrm{sub} = 320^\circ$C and $T_\mathrm{sub} = 380^\circ$C as well as an oxygen pressure of $p = 25\,\mu$bar and $p = 10\,\mu$bar, respectively.\cite{Fischer:2018,Fischer:2020} The (001)-oriented Y$_3$Fe$_5$O$_{12}$ film is deposited with $2.0\,\mathrm{J/cm^2}$ and $f=10$\,Hz on single-crystalline, (001)-oriented Gd$_3$Ga$_5$O$_{12}$ (GGG) substrates at $T_\mathrm{sub} = 450^\circ$C and $p = 25\,\mu$bar.
Without breaking the vacuum, all films are covered $\textit{in-situ}$ by thin Pt layers via electron beam evaporation at room temperature under ultra-high vacuum.

The structural properties of the samples are investigated in detail by high-resolution X-ray diffractometry (HR-XRD). The $2\theta$-$\omega$ scans shown in Fig.~\ref{fig:XRD}  display only reflections from the oxide thin films, the Pt layers, and the respective substrates. No secondary crystalline phases are detected. Finite thickness fringes around the oxide thin film reflections as well as the Pt~(111)-reflection reveal a coherent growth with low surface roughness and high crystalline quality of the respective layers, which is further confirmed by the full width at half maximum (FWHM) of the rocking curves around the symmetric reflections of $<0.03^\circ$. The in-plane orientations and strain states of the oxide layers are investigated by reciprocal space mappings, demonstrating epitaxial relations $[0001]\alpha$-$\mathrm{Fe}_2\mathrm{O}_3 \negthickspace \parallel \negthickspace [0001]\mathrm{Al}_2\mathrm{O}_3$ and $[10\overline{1}0]\alpha$-$\mathrm{Fe}_2\mathrm{O}_3 \negthickspace \parallel \negthickspace [10\overline{1}0]\mathrm{Al}_2\mathrm{O}_3$ as well as $[111]\mathrm{NiO} \negthickspace \parallel \negthickspace [0001]\mathrm{Al}_2\mathrm{O}_3$ and $[110]\mathrm{NiO} \negthickspace \parallel \negthickspace [10\overline{1}0]\mathrm{Al}_2\mathrm{O}_3$. The lattice constants of the $\alpha$-Fe$_2$O$_3$ and NiO thin films are very close to their respective bulk values, indicating a nearly fully relaxed strain state. However, the Y$_3$Fe$_5$O$_{12}$ thin films are fully strained on the GGG substrates due to the low lattice mismatch of only 0.06\%. For further details, we refer the reader to Refs.~[\onlinecite{Fischer:2020}] ($\alpha$-Fe$_2$O$_3$), [\onlinecite{Fischer:2018}] (NiO), or [\onlinecite{Althammer:2013}] (Y$_3$Fe$_5$O$_{12}$).  

\begin{figure}
  \includegraphics[width=0.7\columnwidth]{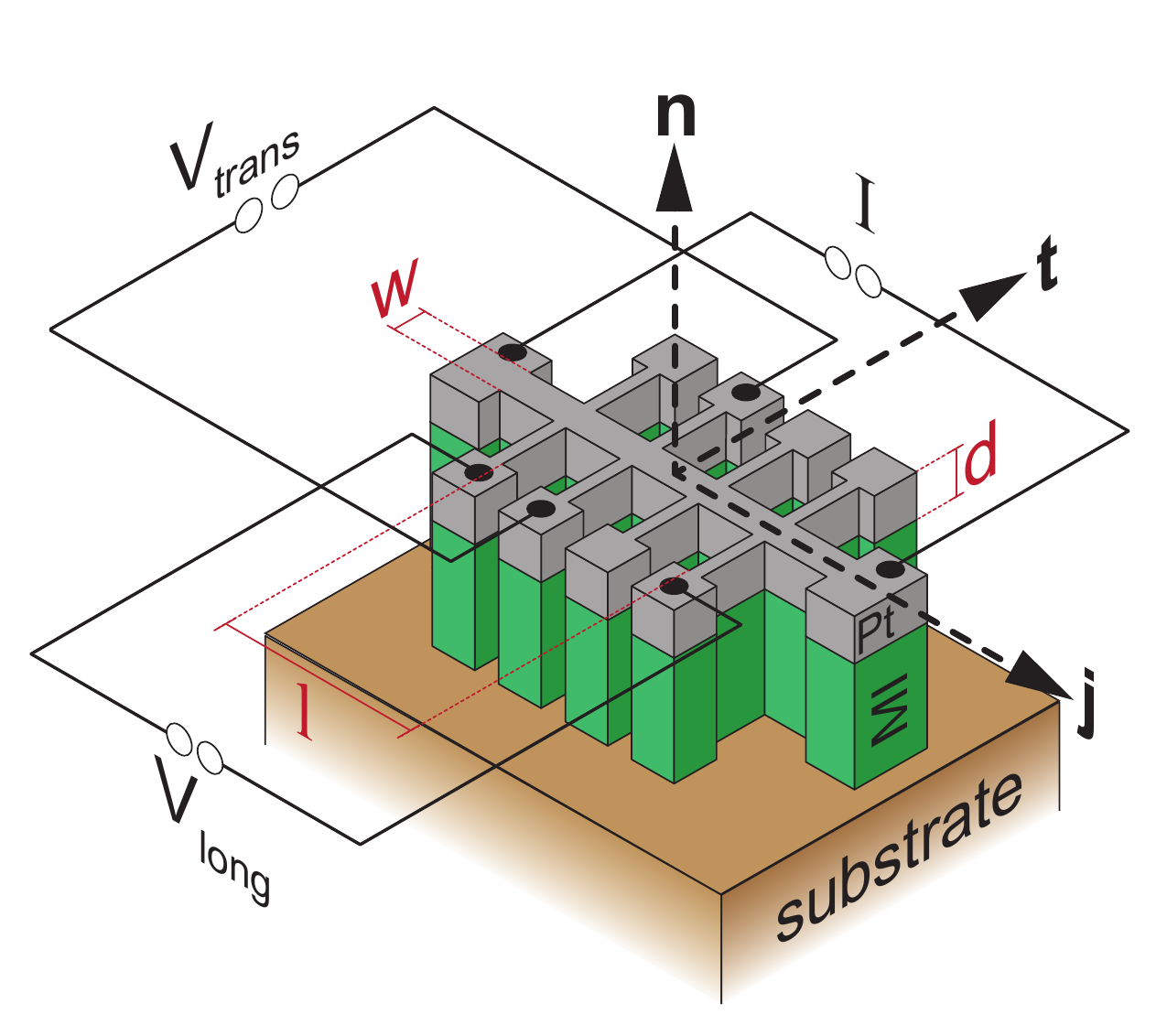}
  \caption{
    \label{fig:Hallbar}
		Measurement geometry for a thin film bilayer, consisting of a magnetic insulator (MI, green) and a metallic Pt-electrode (gray) with thickness $d$, and the coordinate system with the current direction $\mathbf{j}$, the transverse direction $\mathbf{t}$ and the normal direction $\mathbf{n}$. The bilayer is patterned into a Hall bar-shaped mesa structure with width $w$ and length $l$ via photolithography. $\rho_{\mathrm{long}}$ ($\rho_{\mathrm{trans}}$) is determined along $\mathbf{j}$ ($\mathbf{t}$) by measuring the voltage drop $V_{\mathrm{long}}$ ($V_{\mathrm{trans}}$) while applying an electrical current $I$ along $\mathbf{j}$.
    }
\end{figure}

The magnetic properties are investigated via superconducting quantum interference device (SQUID) magnetometry. From room-temperature measurements of the magnetization versus the magnetic field applied in the film plane and after subtracting the diamagnetic (paramagnetic) background of the Al$_2$O$_3$ (GGG) substrate, we determine the saturation magnetization $M_\mathrm{s}$. Within experimental error, we find $M_\mathrm{s} = 10$\,kA/m ($\alpha$-Fe$_2$O$_3$)\cite{Fischer:2020}, $M_\mathrm{s} \simeq 0$ (NiO), and $M_\mathrm{s} = 110$\,kA/m (Y$_3$Fe$_5$O$_{12}$)\cite{Althammer:2013}. All values are in agreement with data reported in the literature.

\subsection{
  \label{sec:expt-ADMR}
  Angle-dependent magnetotransport
  }

Via photolithography and Ar ion milling, Hall bar-shaped mesa structures with a nominal width of $w\,=\,80\,\mu$m and a longitudinal contact separation (length) of $l\,=\,600\,\mu$m are patterned into the bilayers. While a dc current of $I = \pm100\,\mu$A is applied along the $\mathbf{j}$ direction, the longitudinal ($V_\mathrm{long}$) and the transverse ($V_\mathrm{trans}$) voltages are simultaneously measured in a standard four-probe configuration (cf.~Fig.~\ref{fig:Hallbar}). The current-reversal method is applied to eliminate thermoelectric effects.\cite{Ganzhorn:2016} The resistivities are calculated via $\rho_\mathrm{long} = V_\mathrm{long}\;w\;d/(I\;l)$ and $\rho_\mathrm{trans} = V_\mathrm{trans}\;d/I$, where $d$ is the Pt layer thickness. We perform angle-dependent magnetoresistance (ADMR) measurements by rotating an external magnetic field $\mathbf{H}$ of constant magnitude $H$ in the three orthogonal planes ip, oopj, and oopt (see Fig.~\ref{fig:alpha-phi}).

\section{
  \label{sec:results}
  Results and Discussion
  }

For the following discussion, we focus on ADMR measurements on three bilayer samples: A $\alpha$-Fe$_2$O$_3$/Pt (91.4\,nm/3.0\,nm) bilayer, a NiO/Pt (120.0\,nm/3.5\,nm) bilayer, and a Y$_3$Fe$_5$O$_{12}$/Pt (46.1\,nm/3.5\,nm) bilayer. An overview of the investigated samples is given in Table~\ref{tab:samples}. Since the SMR amplitude depends on the thickness of the Pt electrode\cite{Althammer:2013} we here investigate samples with comparable Pt thickness, approximately equal to twice of the spin diffusion length of our Pt which ensures the maximum SMR signal. 

\begin{table}
  \caption{
    \label{tab:samples}
    Overview of the investigated bilayer thin film samples. The layer thickness is denoted by $d$; $T_\mathrm{N}$ and $T_\mathrm{C}$ are the N\'{e}el and Curie temperature, respectively.}
  \begin{ruledtabular}
  \begin{tabular}{cccccc}
    Material & $d$  & Magnetic & $T_\mathrm{N}$, $T_\mathrm{C}$ \\
             & (nm) & order    & (K)                            \\
    \hline
    $\alpha$-Fe$_2$O$_3$/Pt & $91.4$ / $3.0$  & AF\footnotemark[1] & 953 \\
    NiO/Pt                  & $120.0$ / $3.5$ & AF\footnotemark[1] & 523 \\
    Y$_3$Fe$_5$O$_{12}$/Pt  & $46.1$ / $3.5$  & FM\footnotemark[2] & 560 \\
  \end{tabular}
  \footnotetext[1]{antiferromagnetic}
  \footnotetext[2]{ferrimagnetic}
  \end{ruledtabular}
\end{table}

\subsection{
  \label{sec:results-ip}
  In-plane rotations of the magnetic field
  }

\begin{figure}
  \includegraphics[width=0.9\columnwidth]{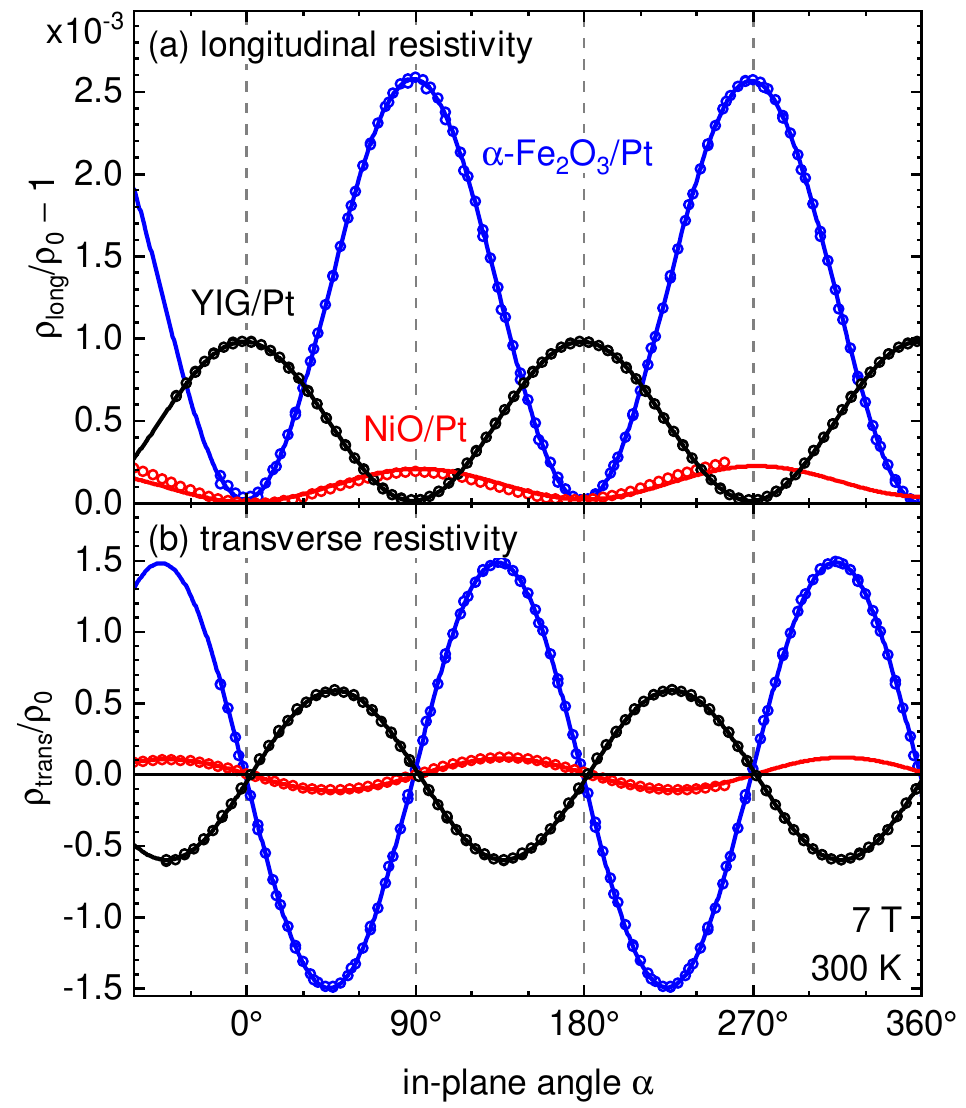}
  \caption{
    \label{fig:ip}
    In-plane ADMR at 300\,K in a magnetic field of 7\,T of antiferromagnetic (0001)-oriented $\alpha$-Fe$_2$O$_3$/Pt (blue) and (111)-oriented NiO/Pt (red) bilayers as well as a ferrimagnetic (001)-oriented Y$_3$Fe$_5$O$_{12}$ (YIG)/Pt bilayer (black). The symbols represent the normalized (a) longitudinal ($\rho_\mathrm{long}$) and (b) transverse ($\rho_\mathrm{trans}$) resistivities measured while rotating the magnetic field in the film plane (ip-rotations). The data is plotted as a function of the magnetic field orientation $\alpha$. The lines are fits to the data using $\cos2\alpha$ and $\sin2\alpha$ functions analogous to Eqs.~(\ref{eq:SMR-FMI}), (\ref{eq:SMR-AFI}).
    }
\end{figure}

When rotating the magnetic field within the film plane (ip-rotations), $\rho_\mathrm{long}(\alpha)$ and $\rho_\mathrm{trans}(\alpha)$ display the characteristic SMR oscillations with $180^\circ$ periodicity (Fig.~\ref{fig:ip}), as expected according to our considerations in section~\ref{sec:SMR-versusM}. The ADMR of the antiferromagnetic $\alpha$-Fe$_2$O$_3$/Pt (blue symbols in Fig.~\ref{fig:ip}) and NiO/Pt (red symbols in Fig.~\ref{fig:ip}) bilayers reveal the same angle dependence, which is, however, shifted by $90^\circ$ relative to that of the ferrimagnetic insulator Y$_3$Fe$_5$O$_{12}$/Pt bilayer (black symbols in Fig.~\ref{fig:ip}). This phase shift by $90^\circ$ represents the characteristic signature of the antiferromagnetic (``negative'') SMR.\cite{Hoogeboom:2017, Fischer:2018, Baldrati:2018} 

The amplitude of the ADMR, however, strikingly differs not only between the antiferromagnetic insulators- and the ferrimagnetic insulator-based heterostructures, but even among the investigated antiferromagnetic insulator bilayers themselves. At a magnetic field magnitude of 7\,T, the ADMR amplitude in $\alpha$-Fe$_2$O$_3$/Pt reaches twice the value of Y$_3$Fe$_5$O$_{12}$/Pt and is by an order of magnitude larger than in NiO/Pt. 

\begin{figure}
  \includegraphics[width=0.9\columnwidth]{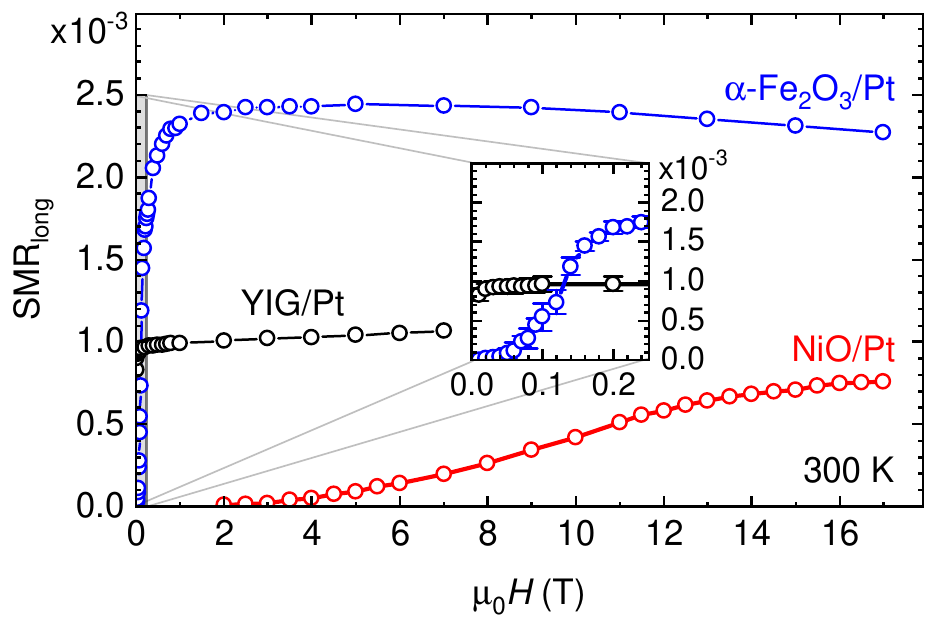}
  \caption{
    \label{fig:SMRlong}
    Comparison of the SMR amplitudes $\mathrm{SMR_{long}}$ of antiferromagnetic (0001)-oriented $\alpha$-Fe$_2$O$_3$/Pt (blue), (111)-oriented NiO/Pt (red), and ferrimagnetic (001)-oriented Y$_3$Fe$_5$O$_{12}$ (YIG)/Pt (black). The data were derived from in-plane longitudinal ADMR measurements (cf. Fig.~\ref{fig:ip}(a)) at 300\,K in different external magnetic fields $H$. The lines are guides to the eye.
    }
\end{figure}

For a quantitative analysis, we fit the experimental data of $\rho_{\mathrm{long}}(\alpha)$ and $\rho_{\mathrm{trans}}(\alpha)$ to $\cos2\alpha$ and $\sin2\alpha$ functions (see Eqs.~(\ref{eq:SMR-FMI}), (\ref{eq:SMR-AFI}) and solid lines in Fig.~\ref{fig:ip}) and determine the amplitude of the ADMR normalized to $\rho_0$, which we plot as a function of the applied magnetic field magnitude $H$ in Fig.~\ref{fig:SMRlong}.
For NiO/Pt, we find a continuous increase of the SMR amplitude $\mathrm{SMR_{long}}$ over the whole field range up to 17\,T (red symbols in Fig.~\ref{fig:SMRlong}). For small fields, $\mathrm{SMR_{long}}$ increases proportionally to $H^2$ and starts to saturate from about 12\,T, as we reported in Ref.~[\onlinecite{Fischer:2018}]. However, for $\alpha$-Fe$_2$O$_3$/Pt, a saturation of $\mathrm{SMR_{long}}$ is already obtained for magnetic fields larger than around 3\,T followed by a gradual decrease from 5\,T to 17\,T.\cite{Fischer:2020} 

We explained this behavior in a comprehensive model, taking into account antiferromagnetic domains within the easy plane of NiO and $\alpha$-Fe$_2$O$_3$ (see Eq.~(\ref{eq:SMR-AFI-domains})). Without applied magnetic field, due to the three fold magnetic anisotropy, NiO and $\alpha$-Fe$_2$O$_3$ develop three types of antiferromagnetic domains labeled by $k=1,2,3$ with equal distribution $\xi_1=\xi_2=\xi_3=1/3$. Since the SMR effect can not distinguish between domains with N\'{e}el vectors $\pm\mathbf{L}=\pm(\mathbf{M}_A-\mathbf{M}_B)/2$, we neglect antiferromagnetic $180^\circ$-domains in the following. A finite magnetic field applied within the easy plane of NiO and $\alpha$-Fe$_2$O$_3$, splits the degeneracy of the energetically equivalent antiferromagnetic domains ($\xi_1\neq\xi_2\neq\xi_3$) and pushes the domain walls towards the energetically unfavorable ones. Increasing the magnetic field magnitude, the fraction of the energetically favorable domains increases until a single antiferromagnetic domain (except for antiferromagnetic $180^\circ$-domains) is present. By minimizing the free energy density taking into account the Zeeman energy and the so called destressing energy with the corresponding destressing field $H_\mathrm{dest}$, which is analog to the demagnetization energy in ferromagnets,\cite{Gomonay:2002, Gomonay:2004} we obtain the domain fraction as a function of the magnetic field magnitude. With Eq.~(\ref{eq:SMR-AFI-domains}), we therefore get the longitudinal ($\rho_\mathrm{long}$) and the transverse ($\rho_\mathrm{trans}$) resistivity within the SMR theory\cite{Fischer:2018}      
\begin{eqnarray}
  \rho_{\mathrm{long}}(\alpha)  &=& \rho_{0} + \frac{\rho_1}{2} \left[1 - \frac{H^2}{H_\mathrm{MD}^2} \cos2\alpha\right] \nonumber \\
	\rho_{\mathrm{trans}}(\alpha)  &=& -\frac{\rho_3}{2} \frac{H^2}{H_\mathrm{MD}^2} \sin2\alpha
  \label{eq:SMR-AFI-domains-HMD}
  \; .
\end{eqnarray}
Here, $H_\mathrm{MD}$ represents the monodomainization field $H_\mathrm{MD}=2\sqrt{H_\mathrm{dest}\;H_\mathrm{ex}}$ with the exchange field $H_\mathrm{ex}$. For magnetic field magnitudes $H$ larger than the monodomainization field $H_\mathrm{MD}$ the antiferromagnetic insulator is in a single domain state (apart from $180^\circ$-domains) and the N\'{e}el vector $\pm\mathbf{L}$ perpendicularly follows the magnetic field leading to the angle-dependence of $\rho_{\mathrm{long}}(\alpha)$ and $\rho_{\mathrm{trans}}(\alpha)$ described by Eq.~(\ref{eq:SMR-AFI}). For magnetic fields $H < H_\mathrm{MD}$, Equation~(\ref{eq:SMR-AFI-domains-HMD}) yields a SMR amplitude of  
\begin{eqnarray}
  \mathrm{SMR}_{\mathrm{long}}  & \approx & \frac{\rho_1}{\rho_0} \frac{H^2}{H_\mathrm{MD}^2}
  \label{eq:SMR-domains-HMD}
  \; 
\end{eqnarray}
with the reasonable assumption $\rho_1 \ll \rho_0$.\cite{Fischer:2018} From this $H^2$-dependence of $\mathrm{SMR}_{\mathrm{long}}$, we can extract the monodomainization fields $H_\mathrm{MD}$ of our NiO and $\alpha$-Fe$_2$O$_3$ thin films to 13.4\,T and 0.24\,T, respectively. This result points to a much lower destressing field $H_\mathrm{dest}$ and therefore to lower magnetoelastic stress fields in $\alpha$-Fe$_2$O$_3$ compared to NiO. This is reasonable, since the magnetostriction $\lambda = 4\times10^{-6}$ in the basal plane of $\alpha$-Fe$_2$O$_3$ at 293\,K\cite{Voskanyan:1968} is by a factor of about 20 smaller than in NiO. This demonstrates that the SMR is a versatile tool to investigate magnetoelastic effects in antiferromagnetic insulator thin film heterostructures.

However, as obvious in Fig.~\ref{fig:SMRlong}, the SMR amplitude of NiO/Pt and $\alpha$-Fe$_2$O$_3$/Pt does not saturate at the expected monodomainization fields $H_\mathrm{MD}$ but still increases up to 17\,T and 3\,T, respectively. This is most likely caused by pinning effects of the magnetic domain walls, which are neglected in the theory above. Therefore, a magnetic field $H_\mathrm{SD}$ much higher than the theoretical monodomainization field $H_\mathrm{MD}$ is required to reach a single domain state in the NiO and $\alpha$-Fe$_2$O$_3$ thin films. Interestingly, in $\alpha$-Fe$_2$O$_3$/Pt bilayers, the SMR amplitude gradually decreases to a magnetic field of 17\,T. This can be traced back to an increasing canting of the antiferromagnetic sublattices, which reduces the projection of the sublattice magnetizations on the $\mathbf{t}$-direction, thereby decreasing the SMR amplitude SMR$_\mathrm{long}$.\cite{Cheng:2019, Fischer:2020} 

Still remarkable is the large maximum SMR amplitude of $2.5\!\times\!10^{-3}$ in antiferromagnetic $\alpha$-Fe$_2$O$_3$/Pt as compared to $1.3\!\times\!10^{-3}$ in ferrimagnetic Y$_3$Fe$_5$O$_{12}$/Pt, although the magnetic sublattices in both materials consist of Fe$^{3+}$ ions. Even though the origin of the observed differences is not yet clear, they may be caused by differences in the interface quality, different values of the Gilbert damping, or different magnetic moment densities, causing different efficiencies for the spin transfer from the heavy metal to the magnetic material.

\subsection{
  \label{sec:results-oop}
  Out-of-plane rotations of the magnetic field
  }

\begin{figure}
  \includegraphics[width=0.9\columnwidth]{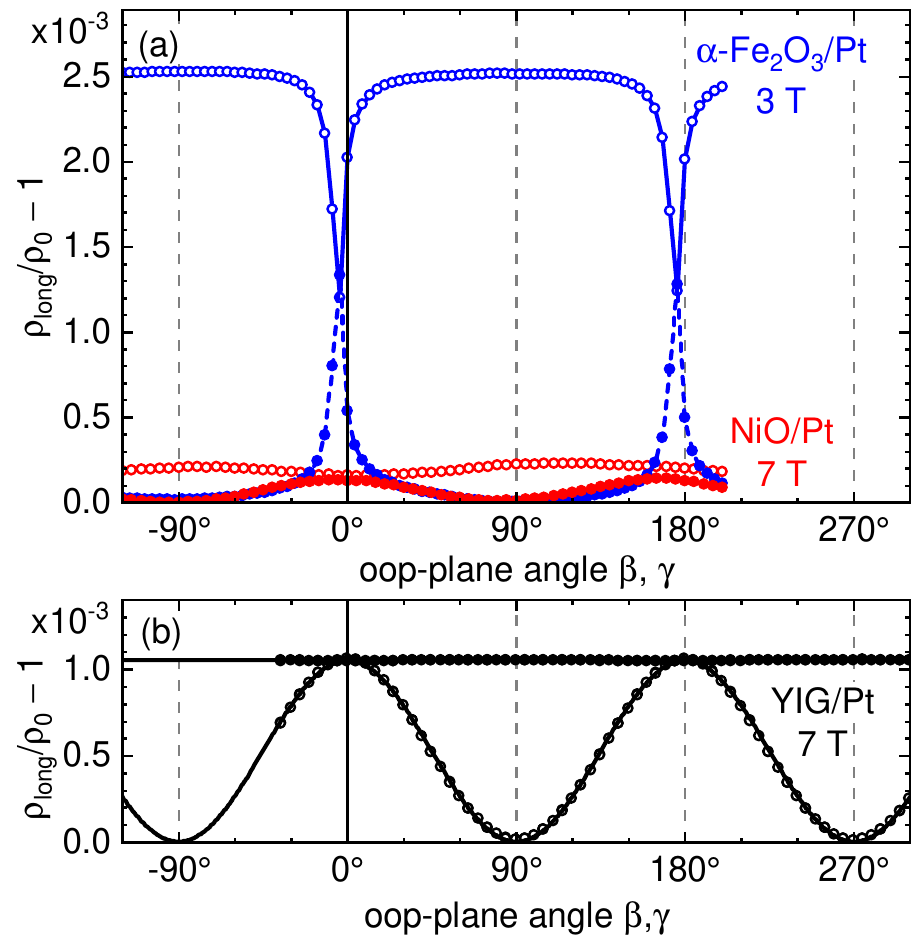}
  \caption{
    \label{fig:oop}
    Out-of-plane ADMR of (a) antiferromagnetic (0001)-oriented $\alpha$-Fe$_2$O$_3$/Pt (blue) and (111)-oriented NiO/Pt (red) bilayers as well as (b) a ferrimagnetic (001)-oriented Y$_3$Fe$_5$O$_{12}$/Pt bilayer at 300\,K. The symbols represent the normalized longitudinal resistivities while rotating the magnetic field out of the film plane orthogonal to the current direction $\mathbf{j}$ (``oopj'', open symbols) or orthogonal to the transverse direction $\mathbf{t}$ (``oopt'', full symbols), respectively. The magnetic field orientation is given with respect to the normal direction $\mathbf{n}$ by the angles $\beta$ or $\gamma$, respectively. The data were taken at 7\,T for  the NiO/Pt and Y$_3$Fe$_5$O$_{12}$/Pt bilayers and at 3\,T for the $\alpha$-Fe$_2$O$_3$/Pt bilayer. For Y$_3$Fe$_5$O$_{12}$/Pt, the black lines are fits to the data according to Eqs.~(\ref{eq:SMR-FMI}).
    }
\end{figure}

For out-of-plane rotations of the magnetic field $\mathbf{H}$, the angle-dependence of the longitudinal resistance $\rho_\mathrm{long}$ of the antiferromagnetic NiO/Pt and $\alpha$-Fe$_2$O$_3$/Pt bilayers is qualitatively different compared to the ADMR of the ferrimagnetic Y$_3$Fe$_5$O$_{12}$/Pt bilayer (cf.~Fig.~\ref{fig:oop}(a) and (b)). The prototype ferrimagnetic Y$_3$Fe$_5$O$_{12}$/Pt bilayer shows a sinusoidal oscillation of $\rho_\mathrm{long}$ in oopj-rotations of the magnetic field (open symbols in Fig.~\ref{fig:oop}(b)) and a nearly constant high resistive state for oopt-magnetic field rotations (full symbols in Fig.~\ref{fig:oop}(b)). This behavior is expected from Eqs.~(\ref{eq:SMR-FMI}) and well established in literature.\cite{Althammer:2013}

For the antiferromagnetic bilayer samples, however, we observe a finite angle-dependence of $\rho_\mathrm{long}$ for both oopj (open symbols in Fig.~\ref{fig:oop}(a)) and oopt (full symbols in Fig.~\ref{fig:oop}(a)) magnetic field rotations. This angle-dependence of $\rho_\mathrm{long}$ can be explained by the projection $H_\mathrm{ip}$ of $\mathbf{H}$ onto the magnetically easy film plane: For $H_\mathrm{ip}=0$, i.e. $\mathbf{H} \parallel \pm\mathbf{n}$ ($\beta,\gamma = 0^\circ$ or $180^\circ$), the antiferromagnetic insulator will form a multidomain state with equal distribution resulting in $\rho_\mathrm{long} = \rho_0 + \rho_1/2$ (cf.~Eq.~(\ref{eq:SMR-AFI-oop})).\cite{Cheng:2019, Fischer:2020} On the other hand, if $H_\mathrm{ip}$ is large enough to form a single-domain state ($H_\mathrm{ip} > H_\mathrm{SD}$), $\rho_\mathrm{long}$ will reach $\rho_0 + \rho_1$ or $\rho_0$ in oopj- or oopt-magnetic field rotations, respectively (cf.~Eqs.~(\ref{eq:SMR-AFI})). However, for magnetic field magnitudes $0 < H < H_\mathrm{SD}$, we expect a continuous modification of the domain fraction within the easy plane, due to the steady increase of $H_\mathrm{ip}$ while rotating the magnetic field $\mathbf{H}$ from $\mathbf{H} \parallel \pm\mathbf{n}$ towards the in-plane $\mathbf{t}$- or $\mathbf{j}$-direction. For oopj- and oopt-magnetic field rotations, $H_\mathrm{ip}$ along the $\mathbf{t}$- and the $\mathbf{j}$-direction is given by $H_\mathrm{ip,t}=H \sin\beta$ and $H_\mathrm{ip,j}=H \sin\gamma$, respectively. Assuming $H_\mathrm{SD}=H_\mathrm{MD}$, i.e. neglecting domain wall pinning or other extrinsic effects, equation~(\ref{eq:SMR-AFI-domains-HMD}) then yields  
\begin{eqnarray}
\mathrm{oopj} &:&\;\rho_\mathrm{long}^\mathrm{AFI}(\beta) \; = \rho_{0} + \frac{\rho_1}{2} \left[1 + \frac{H^2}{H_\mathrm{MD}^2} \sin^2\beta\right] \nonumber \\
\mathrm{oopt} &:&\;\rho_\mathrm{long}^\mathrm{AFI}(\gamma) \; = \rho_{0} + \frac{\rho_1}{2} \left[1 - \frac{H^2}{H_\mathrm{MD}^2} \sin^2\gamma\right] 
     \;\; .
		\label{eq:SMR-AFI_oop_MD}
\end{eqnarray}
We therefore expect a sinusoidal variation of ADMR around $\rho_\mathrm{long}^\mathrm{AFI}(\beta) \; = \rho_{0} + \rho_1/2$ in out-of-plane magnetic field rotations for $H < H_\mathrm{MD}=H_\mathrm{SD}$.

This sinusoidal ADMR is clearly observed in our NiO/Pt bilayer sample at $\mu_0 H=7$\,T$<\mu_0 H_\mathrm{MD}$ (red symbols in Fig.~\ref{fig:oop}(a)). By fits to the out-of-plane ADMR data of NiO/Pt, a mono\-domainization field of $\mu_0 H_\mathrm{MD} = 12.4$\,T can be extracted. This corresponds fairly well with the monodomainization field determined from the magnetic field-dependence of SMR$_\mathrm{long}$ in Fig.~\ref{fig:SMRlong}. However, due to the much smaller monodomainization field $H_\mathrm{MD}$ of $\alpha$-Fe$_2$O$_3$, $H_\mathrm{ip}=H_\mathrm{SD}$ is already reached for $\pm 30^\circ$ away from $\beta,\gamma = 0^\circ$ or $180^\circ$ in out-of-plane ADMR measurements of $\alpha$-Fe$_2$O$_3$/Pt at 3\,T (blue symbols in Fig.~\ref{fig:oop}(a)). This results in two distinct resistivity values of $\rho_\mathrm{long}$, which can be identified with $\rho_0+\rho_1$ and $\rho_0$ (cf.~Eqs.~(\ref{eq:SMR-AFI})). As expected, for $\beta = \gamma = 0^\circ, 180^\circ$, i.e. for $\mathbf{H} \parallel \pm\mathbf{n}$, the in-plane projection $H_\mathrm{ip}$ is zero leading to $\rho_\mathrm{long} = \rho_{0} + \rho_1/2$ (cf.~Eq.~(\ref{eq:SMR-AFI-oop})) and resulting in sharp dip/peak structures.

\begin{figure}
  \includegraphics[width=0.9\columnwidth]{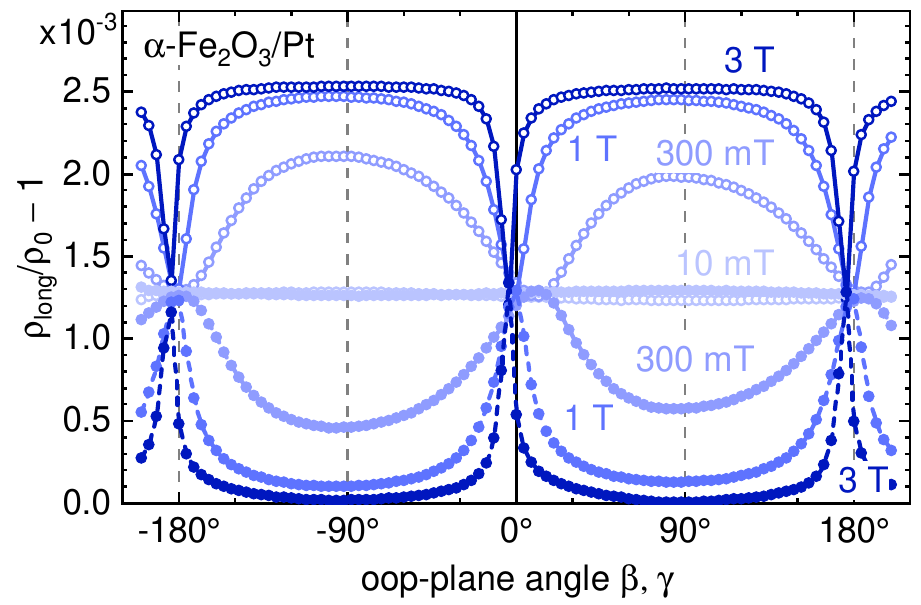}
  \caption{
    \label{fig:FEOS11-oopj-oopt}
    Out-of-plane ADMR of the (0001)-oriented $\alpha$-Fe$_2$O$_3$/Pt bilayer sample at 300\,K in different magnetic field magnitudes $\mathbf{H}$. The open and full symbols represent the normalized longitudinal resistivities while rotating $\mathbf{H}$ out of the film plane orthogonal to the current direction $\mathbf{j}$ (``oopj'') or orthogonal to the transverse direction $\mathbf{t}$ (``oopt''), respectively. The lines are guides to the eye.
    }
\end{figure}

To further prove this scenario, we investigate $\rho_\mathrm{long}$ of the $\alpha$-Fe$_2$O$_3$/Pt bilayer in more detail for oopj- and oopt-rotations at different magnitudes $H$ of the magnetic field (open and full symbols in Fig.~\ref{fig:FEOS11-oopj-oopt}). For low magnetic fields $\mu_0 H=10$\,mT, we observe an almost constant value of $\rho_\mathrm{long} \simeq \rho_0 + \rho_1/2$, indicating a three-domain state regardless of the direction of the external magnetic field $\mathbf{H}$. Increasing the magnetic field magnitude $\mathbf{H}$, the angle-dependence of $\rho_\mathrm{long}$ shows the dip/peak structures described above. However, the expected sinusoidal ADMR for magnetic fields $H < H_\mathrm{MD}$ is only partially observable most probably due to pinning effects of magnetic domains. This might also explain the asymmetric ADMR as well as the finite shift of the minimum/maximum away from $\beta=\gamma=0^\circ, \pm 180^\circ$ for a magnetic field value of 300\,mT. At $\mu_0 H=\mu_0 H_\mathrm{SD}=3$\,T, we observe a constant value of $\rho_\mathrm{long}$ around the in-plane directions $\beta$, $\gamma \approx 90^\circ$, indicating a single-domain state within the film plane.   

\section{
  \label{sec:Summary}
  Summary
  }

We investigate the spin Hall magnetoresistance (SMR) in the easy-plane antiferromagnetic insulators $\alpha$-Fe$_2$O$_3$\cite{Fischer:2020} and NiO,\cite{Fischer:2018} covered with the heavy metal Pt. For rotations of the external magnetic field $\mathbf{H}$ in the easy-plane of the antiferromagnetic insulators, we observe sinusoidal resistivity oscillations in $\rho_\mathrm{long}$ and $\rho_\mathrm{trans}$ phase shifted by $90^\circ$ with respect to the dependence observed for the prototype ferrimagnetic insulator Y$_{3}$Fe$_{5}$O$_{12}$/Pt bilayer. This is clear evidence for the fact that the N\'{e}el vector of the antiferromagnets always stays perpendicular to the field direction and follows the external magnetic field $\mathbf{H}$ for magnetic field magnitudes larger than the single domain field $H_\mathrm{SD}$. For lower fields, a multidomain state within the easy plane of the antiferromagnetic insulator is present with domain fractions depending on the direction and the magnitude of the external magnetic field $\mathbf{H}$. This results in a $H^2$-dependence of the SMR-amplitude.\cite{Fischer:2018,Fischer:2020}     

For out-of-plane rotations of $\mathbf{H}$, we find a situation different from the established behavior in ferrimagnetic insulator/heavy metal bilayers. While a sinusoidal oscillation and constant value of $\rho_\mathrm{long}$ in ferrimagnetic insulator Y$_{3}$Fe$_{5}$O$_{12}$/Pt bilayers are observed in oopj- and oopt-magnetic field rotations, respectively, a peak/dip angle-dependence of $\rho_\mathrm{long}$ is observed in $\alpha$-Fe$_2$O$_3$/Pt and NiO/Pt bilayers. This indicates that for high magnetic fields $H > H_\mathrm{SD}$ and magnetic field orientations $\mathbf{H} \nparallel \pm\mathbf{n}$, the antiferromagnetic insulator is single-domain and $\rho_\mathrm{long}$ stays in a high or low resistive state for rotations in the oopj- or oopt-geometry. However, for magnetic fields pointing (anti-)parallel to the surface normal ($\mathbf{H} \parallel \pm\mathbf{n}$), the antiferromagnetic insulator becomes multi-domain with equal distribution of the domains, resulting in a medium resistive state.   

In summary, we are able to provide a comprehensive picture of the spin Hall magnetoresistance (SMR) in antiferromagnetic insulator/heavy metal thin film bilayer heterostructures. We show that, the SMR provides not only information about the orientation of the N\'{e}el vector and the average domain structure at antiferromagnetic insulator/heavy metal interfaces, but also on magnetoelastic effects of the antiferromagnetic layer elastically clamped on a respective substrate. Since the SMR effect can not distinguish between $180^\circ$-domains, new techniques like NV-center microscopy\cite{Gross:2017} are necessary to image antiferromagnetic spin textures in real space. However, since the SMR is a comparably simple method and also applicable at high magnetic fields, it could become a valuable tool for reading out magnetization states in the emerging field of antiferromagnetic spintronics.

\section*{AUTHOR'S CONTRIBUTIONS}
S.~Gepr\"ags and M.~Opel contributed equally to this work. 

\begin{acknowledgments}
We thank Thomas Brenninger, Astrid Habel, and Nynke Vlietstra for technical support. We gratefully acknowledge financial support of the German Research Foundation via Germany's  Excellence Strategy (EXC-2111-390814868). OG acknowledges support from the Alexander von Humboldt Foundation, EU FET Open RIA Grant no. 766566, and DFG (project SHARP 397322108).
\end{acknowledgments}

\section*{DATA AVAILABILITY}
The data that support the findings of this study are available from the corresponding author upon reasonable request.

\providecommand{\noopsort}[1]{}\providecommand{\singleletter}[1]{#1}%

\end{document}